# *$F^2$- Rules for Qualification of Developing and Managing Software Product Line

F. Ahmed

Electrical & Computer Engineering
University of Western Ontario
London Ontario, Canada, N5A5B9
sgraha5@uwo.ca

L.F. Capretz

Electrical & Computer Engineering
University of Western Ontario
London Ontario, Canada, N5A5B9
lcapretz@eng.uwo.ca

**Abstract:** Software product line has emerged as a valuable concept of developing software based on reusable software assets. The concept aims on effective utilization of software assets, reduced time to delivery, improved quality and better benefits to cost ratio of products. In this paper we have defined certain rules for the qualification of developing and managing a software product line. An organization must follow these rules in order to establish and manage software product line effectively.

## 1. Introduction

### 1.1 Problem Definition

A software product line is a set of software-intensive systems sharing a common, managed set of features that satisfy the specific needs of a particular market segment or mission and that are developed from a common set of core assets in a prescribed way [1]. The concept of software product line has become an attractive phenomenon within organizations dealing with software development process. The inception, elaboration and construction phase of software product line requires a careful strategy to implement this concept. Organizations trying to incorporate this concept require certain rules to be followed for effective development and management of software product line. There is a need of defining and summarizing all the necessary guidelines and principles for software product line development and management activities as rules so that they should be strictly and religiously followed for successful outcome of efforts. $F^2$ - Rules covers all the possible aspects of developing and managing a software product line in an organization effectively.

## 2. $F^2$-Rules

The general structures of these rules are "statement portion" and "discussion portion", the statement defines the rule and discussion elaborates the concept of the rule with implications of not following it. Rules are categorized as Core Asset Development Rules, Product Development Rules and Management Rules, and cover all the essential activities of software product line development and management. All the rules are given unique number but it is not required that in the same order these rules should be implemented or followed in the organization.

* The rules are pronounced as F-Square Rules, as the first name of both the authors starts with F, i.e., Faheem and Fernando.

## 2.1 Core Asset Development Rules

- **Rule # 1**

**Statement:** "All the core assets within software product line repository and resulted products must be consistent with the scope of software product line"

**Discussion:**

It is important to define the proper scope of the software product lines, as it is the very basis for the strategic development of product lines [2]. Product line scoping is used to define the product line. Namely, it determines the products that comprise the product line [3]. Once the scope of software product line is defined it is necessary that all the core assets must be consistent with the scope of product line because it is not required to collect all the assets and develop a repository for product line, the aim is to develop core assets for the product line within the scope of the product line so that they can be utilized while developing products. The products developed should fall within the scope of the product line as well.

- **Rule # 2**

**Statement:** "Every component present in the core assets repository must define the variability mechanism to tailor them for effective utilization"

**Discussion:**

Instead of developing and deploying a "fixed" one-of-kind system, it is now common to develop a family of systems whose members differ with respect to functionality or technical facilities offered [4]. Fitting the component into the product without tailoring it is the easiest task, but some time we need to make certain changes in the components to meet the requirements for a particular product. Every component present in the core assets must clearly define the variability mechanism to be used in order to tailor them for reusing. A separate document must be attached with the components, which elaborates this activity.

- **Rule # 3**

**Statement:** "Core assets repository update constantly by adding new assets as product line progress"

**Discussion:**

If we use proactive approach to develop software product line then initially all the core assets are identified and as we progress further products resulted from the product line tends to develop new core assets which must be added into repository so that can be reused for next products. If we use active approach to develop product line then we start developing products and core assets generated during development process constitute the core asset repository. So whatever the approach we adopt to develop software product line, core asset repository should be dynamic and should keep on increasing its size by adding more components.

- **Rule # 4**

**Statement:** "All the COTS present or added into core asset repository must satisfy the cost benefits ratio for the organization"

**Discussion:**

The use of COTS element can result in reduced development cost, development and integration risk and development time [5]. Developing software with as much COTS functionality as possible saves you from reinventing the wheel [6]. But at the same time it is required that the benefit to cost ratio must satisfy the organizational goal otherwise it will considerably increase the overall cost of the product. To achieve the target of product development within budget all the COTS added into the core asset must meet the benefit to cost ratio criteria for the organization.

- **Rule # 5**

**Statement:** "A version control management system should keep track of the core asset development and utilization history"

**Discussion:**

The core asset in the repository are to be used in various products and their versions, it is necessary to keep track of the history of utilization of individual core assets in different products, this history should clearly describe the functionalities used, interface requirements, and any modification done to accommodate the core asset into a new product. And if any considerable modification is done then it should be termed as a new version of the same core asset and ultimately added to the core asset repository with an associated definition of its parent. A comprehensive version control management system of core assets must support the core asset development activity to keep track of all the components and their offspring.

## 2.2 Product Development Rules

- **Rule # 6**

**Statement:** "All the products within the software product line must share a common architecture"

**Discussion:**

Software architecture describes the overall organization of a software system in terms of its constituent elements, including computational units and their interrelationships [7]. Design for reuse and open architectures are of the utmost importance with respect to software architectures for product lines [8]. The purpose of the software product line is not just the reuse, it targets on the effective delivery of shared architecture products. All the products must share a common architecture so that they can be termed as family of products, because a software product line is based on a system family architecture offering a " common set of core assets" [9].

- **Rule # 7**

**Statement:**" A variation among products should remain within the scope of software product line"

**Discussion:**

The products from the software product line may vary from each other's in quality, reliability, functionality, performance etc, but as they share the common architecture so the variation should not be that much high so that they become out from the scope of the product line. Those variations must be handled systematically to accommodate changes in various versions of the product.

- **Rule # 8**

**Statement:** "Every product released from product line should be a valid business case for the organization"

**Discussion:**

The business cases define the marketing strategy of the organization. It explores the market for the profitable business. Every organization always identifies potential business cases in order to capture the market. It is necessary that each products released from the software product line must be a valid business case for the organization so that the organization can ultimately achieve its financial goal along with the justification of the product itself.

- **Rule # 9**

Statement: "Software product line must capable of producing considerable number of products, at least more than one"

**Discussion:**

The main aim behind software product line is to develop a stream of products out of core assets. If the product line is just aimed to produce only one single product then the activity can be regarded as "Just Component Based Development" not the software product line. Therefore the scope and structure of the product line should be to develop all those products that meet the business case criteria for the organization.

- **Rule # 10**

**Statement:** "Every product released from the software product line must meet the qualification criteria of the organization"

**Discussion:**

Every organization defines their parameters for the qualification of a product along with the standard acceptance criteria. A product is only feasible if it meets the qualification criteria as defined earlier to its development. Therefore it must be clearly defined what are the qualification criteria of the software product line so that all the products resulted from the development of software product line must meet those criteria.

### 2.3 Management Rules

- **Rule # 11**

**Statement:** "A multi dimensional configuration management approach should handle the configuration management issues present in the software product line "

**Discussion:**

The configuration management issues are imperative in software product lines as it deals with number of resulted products with different versions and releases as well as numerous number of core assets with different versions. Therefore a multi dimensional approach of configuration management should be adopted to cope up with the issue. Such an environment may be defined as configuration management of configuration management system. In this approach a separate configuration management systems are applied to product and core assets and on the top of those two-configuration management another configuration management handle the coordinated issues of both.

- **Rule # 12**

**Statement:** "A comprehensive description and analysis of domain should be performed for which the software product line is to be developed"

**Discussion:**

Domain analysis is the systems engineering of a family of systems in an application domain through development and application of reusable assets [10]. Domain analysis entails developing a complete and rigorous domain model and associated generic architecture as a precursor to developing a set of reusable components for repeated application in developing systems in the domain [10]. The domain analysis for software product line will support the development as well as reusing the core assets in development It gives a guideline for identification of potential core assets for the software product line and provides a basis for the business case evaluation of products.

- **Rule # 13**

**Statement:** "The ROI (Return on Investment) of the software product line must meet the organizational financial goal"

**Discussion:**

The construction of software product line will only be beneficial in term of finance if the organizational ROI meets the out come of software product line. The investment incurred on the product line must justify itself. It is generally incurred that the return of software product line heavily based on the resulted products and gradually increased as the increase in number of products to be delivered to the market.

- **Rule # 14**

**Statement:** "Requirements of the software product line must be clearly defined, analyzed, specified, verified and managed"

**Discussion:**

The manage requirements describes how to elicit, organize, and document required functionality and constraints; track and document tradeoffs and decisions; and easily capture and communicate business requirements [11]. If we perform good requirement management for the software product line then it will help in understanding scope and boundaries of the products to be developed, which ultimately helps us in identifying core assets for the software product line.

- **Rule # 15**

**Statement:** "Requirements of the software product line must define the fundamental products and their features within the product line"

**Discussion:**

Product line requirements define the products and the features of the products in the product line [12]. The requirement engineering of software product line must yield the features of fundamental product. It should describe the core functionality the products are supposed to provide, the properties it must exhibit and associated constraints and quality parameter. The product line requirements also elaborate the variability issue among the products.

- **Rule # 16**

**Statement:** "Organization structure must support the software product line concepts and principles"

**Discussion:**

Software product line approach is little bit different from traditional software development approach. It requires a considerable visualizing approach, management enforcement, communications, discussions and elaborations of what you have and what you can do. Therefore organizational structure must support the concepts and principles of software product line. There should be a clear definition of team and its associated task and duties.

- **Rule # 17**

**Statement:** "All the three essential activities of software product line development must be performed iteratively"

**Discussion:**

The essential activities core asset development, management and product development performed during product line development are linked together are highly iterative. Core assets are used to develop new products and there is a constant chance of adding up the piles of core assets either as an out come of new product development or COTS. The management takes its inputs from core assets and development phase and continuously gives feed back to both. The iterative development approach will allow interacting these three essential activities with each other's to support them.

## 3. Conclusion and Future Research

Software product line development activity is a systematic approach to develop set of products having similarities based on their core functionalities and architecture by reusing what the organization currently have or intend to have. The $F^2$-Rules will enable organizations to build their strategy for product line development as well as to monitor it. It summarized the fundamental and essential procedures and processes to be adopted for developing and managing a software product line effectively. It also helps people to understand what the software product line is and how it is populated with various integral concepts and principles. The rules can be implemented in any order disrespect of their sequence.

Our future research work concentrates on using $F^2$-Rules to evaluate the capability of an organization to develop and manage a software product line. This quantitative measurement will enable an organization to understand the deficiencies in their process and provides a framework to improve the process. We are working on $F^2$-Rules to generate a cost evaluation model for software product line in which these rules will be treated as cost drivers for the economic model.